Mapping tidal flat topography using time-series Sentinel-2 images and ICESat-2 data: A case study in Cixi City

Xiucheng Zheng[a,b,c], Bin Zhou[a,b,d,*], Hui Lei[a,b,c], Qianqian Su[a,b,c] and Yuxuan Jin[c]

a. Institute of Remote Sensing and Earth Science, Hangzhou Normal University, Hangzhou 311121, China
b. Zhejiang Provincial Key Laboratory of Urban Wetlands and Regional Change, Hangzhou 311121, China
c. School of Information Science and Technology, Hangzhou Normal University, Hangzhou 311121, China
d. School of Engineering, Hangzhou Normal University, Hangzhou 311121, China

**Abstract:**
Tidal flat topography provides crucial insights for understanding tidal flats and their dynamic evolution. However, the wide-ranging and rapidly changing nature of tidal flats, which are periodically submerged in shallow water, pose challenges for many current monitoring methods in terms of both efficiency and precision. In this study, we considered the dynamic process of tidal flat submergence and utilized time-series Sentinel-2 images on Google Earth Engine (GEE) to calculate the tidal flat exposure frequency. This information was used to determine the spatial extent of the tidal flats, and subsequently, by employing ICESat-2 data, we established a 1D-linear regression model based on elevation and frequency values, which realizes the inversion of the tidal flat elevation within Cixi City. The study shows the following: (1) the tidal flat exposure frequency and ICESat-2 elevation data exhibit a strong positive correlation ($R^2$=0.85); (2) the tidal flat area within Cixi City is 115.81 $km^2$, and the overall accuracy is 95.36%; and (3) the elevation range of the tidal flats in the study area is between -0.42 and 2.73 m, and the mean absolute error (MAE) is 0.24 m. Additionally, we consider that the temporal resolution of remote sensing imagery plays a crucial role in determining the accuracy of the elevation inversion, and we found that higher tidal flats exhibit better inversion accuracy than lower tidal flats.

**Introduction**

Tidal flats, referring to beaches, rock flats, or mudflats that become exposed in the intertidal zone during low tides and submerged during high tides (Healy et al., 2002), support a remarkably high biomass volume despite relatively low species abundance. They nurture a plethora of invertebrates and provide 10-20 times more aquatic resources compared to deeper coastal waters, supporting abundant marine life and serving as feeding grounds for migratory birds (Reise et al., 2010). Additionally, due to the high sediment carbon storage, estimated to be 50 times higher than terrestrial carbon sinks (Einsele et al., 2001), tidal flats serve as significant carbon sinks. Therefore, in addition to their diverse ecological functions, tidal flats continually generate economic value for society.

Tidal flat topography provides crucial insights for understanding tidal flats and their dynamic evolution(Li et al., 2022a).. Monitoring tidal flat topography contributes a vital role in guiding coastal conservation and development, with spatial distribution and elevation data being key parameters. Traditional field-based research methods can achieve a high level of precision, ideally reaching a 5 cm vertical accuracy for profiles and a 1 cm accuracy for control points (Gorman et al., 1998). However, the extensive and rapidly changing nature of tidal flats, which experience periodic shallow water submergence, presents significant complexity and potential hazards for traditional field surveys. In contrast, remote sensing technology offers an alternative or complementary approach, whether using ship-based or land-based platforms.

Currently, remote sensing techniques used for monitoring tidal flats include synthetic aperture radar (SAR) (Kim et al., 2014), interferometric synthetic aperture radar (InSAR) (Xie et al., 2015), light detection and ranging (LiDAR) (Luo et al., 2017), structure-from-motion (SfM) photogrammetry (Kalacska et al., 2017), and the waterline method (Mason et al., 2001), among others. While enabling the construction of tidal flat digital elevation models (DEMs), SAR, InSAR and LiDAR techniques tend to be costly; however, UAV-based SfM photogrammetry offers cost-effective and accurate mapping of tidal flat topography, albeit within a limited coverage area. The waterline method, proposed by Mason et al. (2001), treats the waterlines as contour lines and utilizes multiple waterline images from different periods to generate a tidal flat DEM through interpolation. Despite requiring significant effort due to the large volumes of data, the advent of the Google Earth Engine (GEE) remote sensing cloud platform has addressed issues related to manual labour and data size. Utilizing the GEE, Chang et al. (2022) calculated the maximum and minimum Normalized Difference Water Index (NDWI) values of time-series images at specific locations. The synthetic images generated from the NDWI maxima and minima were then used by Chang et al. to delineate tidal flat boundaries and achieve highly accurate mapping of tidal flat areas along the Bohai and Yellow Seas. Building on previous research, this study integrates multi-temporal optical image data with LiDAR data from Sentinel-2 and ICESat-2, proposing a tidal range extraction and elevation inversion method based on tidal flat exposure frequency. This approach enables the delineation of the spatial distribution and elevation of tidal flats in the study area and provides as a technical reference for the protection of tidal flats.

**1. Study area and data sources**

1.1 Study area

This study focuses on Cixi City, located on the south shore of Hangzhou Bay, as the study area. The wetlands in Hangzhou Bay feature complex habitats and abundant biodiversity, serving as a crucial ecological barrier in the region. They have profound significance in providing habitats for rare animals, protecting biodiversity, and maintaining regional ecological balance. Due to the distinctive spatial morphology of Hangzhou Bay, a substantial amount of sediment has accumulated along the north convex arc of the south coast within Cixi City, forming an expansive tidal flat landscape.

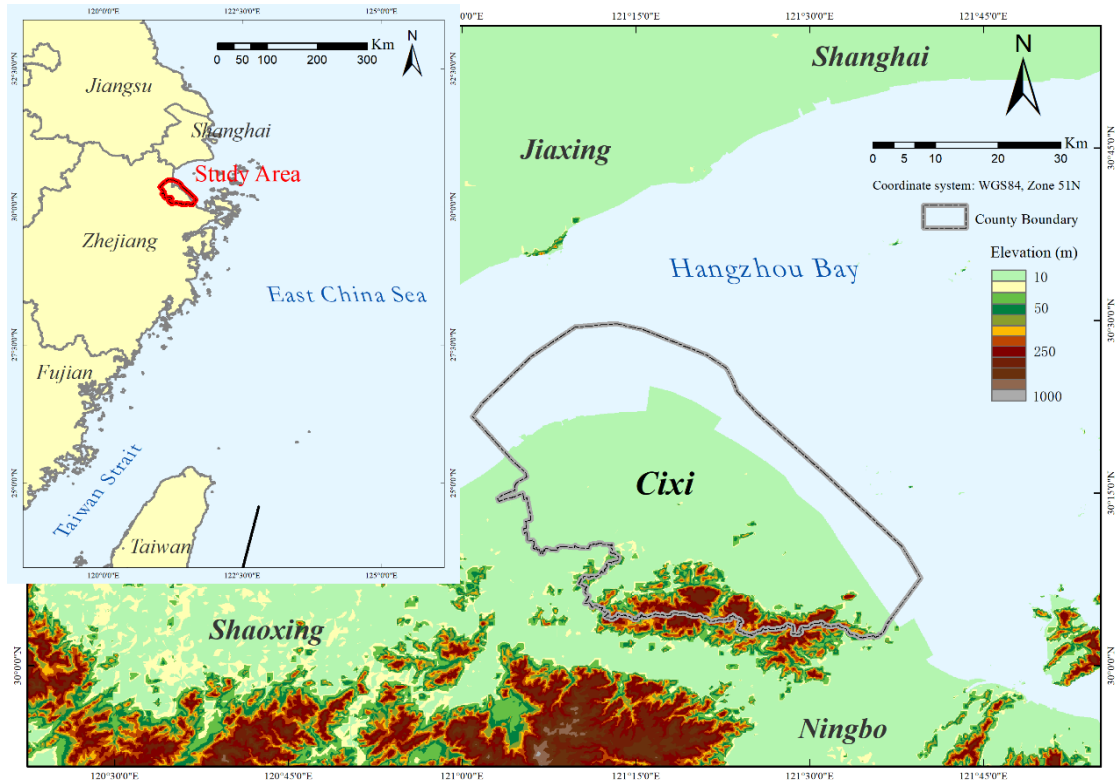

**Fig. 1.** Study area.

1.2 Data sources

The data utilized in this study consist of Sentinel-2 MSI images, ICESat-2 laser altimetry data, and vector boundary data for administrative districts from Sentinel-2 data.

The Sentinel-2 data were obtained from the GEE, specifically the Sentinel-2 surface reflectance dataset in the Earth Engine Data Catalog. The time frame selected for the analysis spans from January 1, 2021, to December 31, 2022. To ensure high-quality imagery, images that had cloud cover exceeding 20% were filtered out, and a cloud mask module was incorporated. A total of 92 scenes were utilized in the computational process.

ICESat-2, officially known as Ice, Cloud and Land Elevation Satellite-2, carries the Advanced Topographic Laser Altimeter System (ATLAS), which emits six 532 nm laser pulses (arranged in 3 pairs) towards the Earth's surface. Each pair consists of a "strong" beam and a "weak" beam, with an energy ratio of approximately 4:1. The beams are spaced approximately 90 m apart within a pair and approximately 3 km apart between pairs. By calculating the time taken for the photons to travel to the Earth and back, the elevation relative to the WGS84 reference ellipsoid is determined(Neumann et al., 2020). Research has demonstrated the outstanding performance of ATLAS, achieving elevation measurement accuracy at the centimetre level (Markus et al., 2017; Parrish et al., 2019). In this study, the Level-2 product ATL03, which incorporates multiple geophysical corrections, was utilized. The data were obtained from the NASA Earth Data portal (https://search.earthdata.nasa.gov/). Specifically, ICESat-2 gt2r data from March 21, 2021, were utilized for inversion, and ICESat-2 gt2l data from November 20, 2021, were used for validation.

The administrative boundary data were acquired from the Resource and Environment Science and Data Center of the Chinese Academy of Sciences, available at https://www.resdc.cn/.

## 2. Methods

2.1 Water index calculation

Calculating water indices is a straightforward method that enables rapid extraction of water information over a large area (Allen et al., 2018) . Commonly used water indices include the NDWI (McFeeters, 1996) , Modified-NDWI (MNDWI) (Xu, 2006), Automated Water Extraction Index (AWEI) (Feyisa et al., 2014) , and Multi-Band Water Index (MBWI) (Wang et al., 2018) . Among these, the MNDWI index has demonstrated higher accuracy and greater stability (James et al., 2021; Ji et al., 2009; Pekel et al., 2016). Therefore, in this study, the MNDWI index was utilized for water segmentation calculation, as shown in the following equation:

$$MNDWI = \frac{\rho_{\text{GREEN}} - \rho_{\text{SWIR}}}{\rho_{\text{GREEN}} + \rho_{\text{SWIR}}}$$

2.2 Otsu algorithm threshold segmentation

The Otsu algorithm, proposed by Japanese scholar Otsu in 1979 (Otsu, 1979), is a method to establish the threshold for image binary segmentation. In principle, the threshold derived by this algorithm maximizes the interclass variance between the foreground and background regions; thus, it is also known as the maximum interclass variance method. By writing code in the GEE, it is feasible to automatically obtain the threshold value of each greyscale image calculated using the water index and to then perform binary segmentation.

2.3 Calculation of non-water frequency per pixel

Thanks to the convenience afforded by remote sensing cloud computing, concepts such as flooding frequency have arisen in recent research (Li et al., 2022b). The principle behind the flooding frequency concept is to calculate the ratio of the number of times a pixel location is identified as a water pixel to the total number of observations. Similarly, in this study, the tidal flat exposure frequency refers to the proportion of non-water pixels at every pixel location. The calculation formula is as follows:

$$F_E = \frac{N_{TF}}{N_O}$$

where $N_{TF}$ is the number of times the pixel is observed as non-water during the study period, and $N_O$ is the total number of valid observations.

2.4 Elevation inversion

The ICESat-2 ATL03 data were converted into tables utilizing PhoREAL v3.30 software. Based on latitude and longitude data, elevation point data were generated in ArcMap 10.2. A total of 1280 elevation points with a "high" confidence level (signal confidence value of 4) were selected within the spatial extent of the study area. Due to the dense distribution of the ICESat-2 laser foot points, to ensure consistency in scale, we took the average of the elevation points within a frequency pixel as the elevation values. Subsequently, a one-dimensional linear regression model was established based on the elevation and frequency values.

2.5 Accuracy validation

This study performs accuracy validation from two aspects: spatial coverage and tidal flat elevations.

2.5.1 Accuracy verification of the tidal flat spatial coverage

To ensure reasonable and evenly distributed sampling points, the tidal flat range extracted within the region is delineated 2 km outwards as the buffer area for accuracy verification. Several points are randomly generated within the buffer area as sampling points, with a minimum spacing of 100 m between every sampling point. Utilizing remote sensing images captured during moments of high and low tide in the study area, along with historical images from Google Earth and aerial photographs taken by the DJI Phantom 4 Pro UAV, visual interpretation is performed to classify the observation points and sampling points into two classes: water and non-water. Finally, an accuracy evaluation of tidal flat extraction is conducted by formulating a confusion matrix and calculating Kappa coefficients and F1_Score.

2.5.2 Accuracy verification of the tidal flat elevation

In this study, the elevation information of the tidal flats was acquired through UAV-photogrammetry using the DJI Phantom 4 Pro and a handheld RTK during low tide. A tidal flat DEM was constructed, and numerous ICESat-2 laser foot points, which are distributed in strip patterns, were utilized as sample points. The inverted elevation values and the in situ elevation values were extracted and stored in an elevation point data table. The R-squared, root mean square error (RMSE), and mean absolute error (MAE) between the inverted and in situ elevation values were computed.

$$RMSE = \sum_{n=1}^{n=i} \sqrt{\frac{[(X_i)^{insitu} - (X_i)^{inverted}]^2}{i}}$$

$$MAE = \frac{1}{i}\sum_{n=1}^{n=i} |(X_i)^{insitu} - (X_i)^{inverted}|$$

where $(X_i)^{insitu}$ is the measured value and $(X_i)^{inverted}$ is the predicted value.

Additionally, due to the limitations in the UAV range and potential errors in the topographic observations based on oblique photography, another set of ICESat-2 data was directly utilized to validate the elevation of another tidal flat region in the study. Likewise, the scales were unified as in the inversion process.

## 3. Results

3.1 Inversion model

An inversion model was developed based on ICESat-2 elevation values and tidal flat exposure frequency values. The model is represented by a one-dimensional linear regression equation:

$$y = 3.3517x - 0.5199$$

where x is the tidal flat exposure frequency, y is the elevation observed by ICESat-2, and the correlation coefficient $R^2 = 0.85$. This model was utilized to invert the frequency values into elevation values within the study area (Fig. 2).

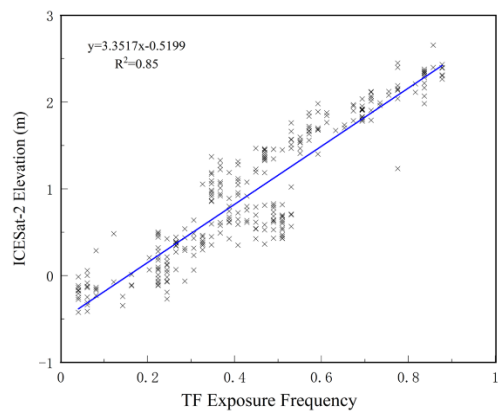

**Fig. 2.** Scatter plots for the establishment of the elevation inversion model.

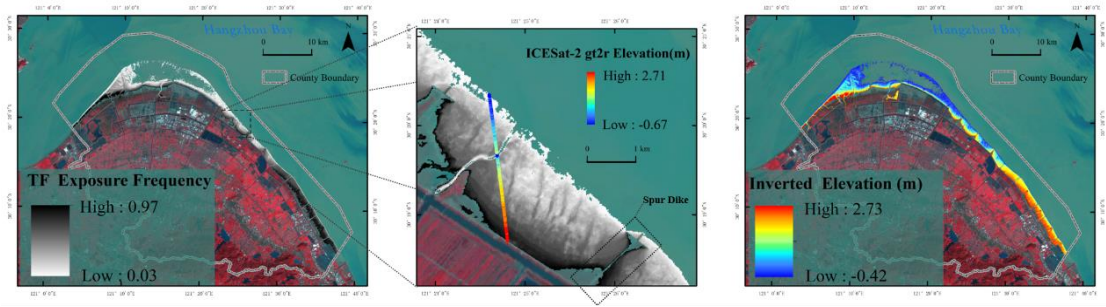

**Fig. 3.** Inverting tidal flat exposure frequency values into elevation values using the elevation inversion model. Basemap: Sentinel-2 annual mean composite image for 2021, R: G: B = Band 8: Band 4: Band 3. Subsequent content also uses this image as the Basemap.

3.2 Spatial and elevation distribution of tidal flats

    Results from statistical analysis reveals the total area of tidal flats within Cixi City is 115.81 km$^2$, with elevations ranging from -0.42 m to 2.73 m. Data in **Fig. 4** illustrates the spatial distribution and elevation range of tidal flats in the Cixi city region along the southern coast of Hangzhou Bay. Section W features long and narrow tidal flats situated at a considerable distance from the shoreline. These flats exhibit a wide distribution of tidal channels on the landward side, with relatively steep slopes but smooth surfaces. In contrast, Section N consists of expansive and predominantly flat tidal flats, with most of them at or below an elevation of 1 m. Facing Hangzhou Bay, this section forms a large opening. Similar to Section W, it also possesses a substantial supratidal zone and numerous tidal channels on the landward side. The tidal flats along the edges of Sections E1 and E2 are aligned parallel to the top of the spur dike. However, there are notable distinctions between these sections. In Section E1, the elevation of the tidal flats is significantly influenced by the spur dike. The elevation positively correlates with the distance to the spur dike and the embankment, resulting in three sides with a higher elevation and one side with a lower elevation. The contour lines form a "horseshoe" distribution pattern. On the other hand, in Section E2, the elevation of tidal flats appears unaffected by the spur dike. The tidal flats exhibit a higher overall elevation and alignment parallel to the shoreline.

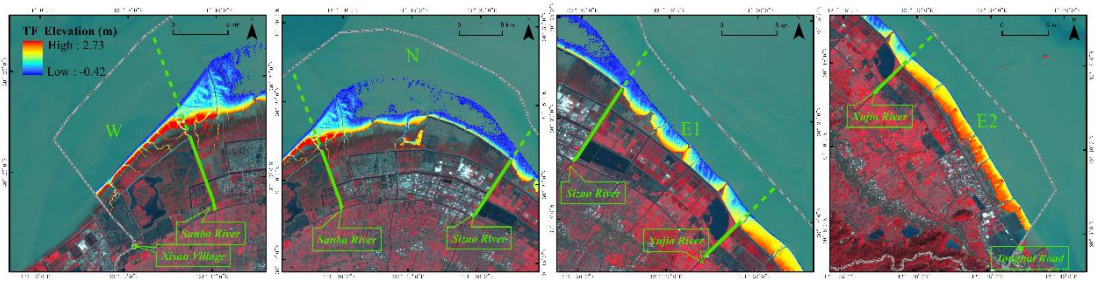

**Fig. 4.** Zoning within the study area and inversion results within each zone. Based on the variations in the spatial distribution and topography of the tidal flats, the study area was divided into four sections: Cixi West (Xisan Village) - Sanba River, Sanba River - Sizao River, Sizao River - Xujia River, and Xujia River - Cixi East (Tonghai Road). For convenience, these sections are referred to as W, N, E1, and E2, respectively.

3.3 Accuracy Verification

3.3.1 UAV-based accuracy verification

    For the purpose of tidal flat spatial accuracy verification, 500 random sampling points were generated within the buffer area. Additionally, 961 photos were captured through UAV aerial photography to construct a tidal flat DEM for tidal flat elevation accuracy verification. Among these

photos, 231 were also utilized for tidal flat spatial accuracy verification (**Fig. 5**).

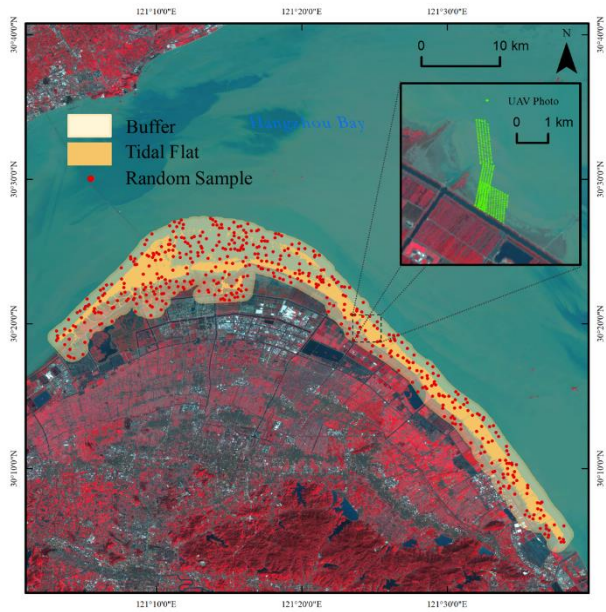

**Fig. 5.** Map of the study area used for accuracy verification, where the white area is the tidal flats, the buffer zone with a 2 km radius is inside the white line, and the green dots in the top-right submap are the UAV aerial photography points.

The data in Table 1 shows the overall accuracy of the spatial extent of the tidal flats is 95.36%, the Kappa coefficient is 0.9069, and the F1_Score is 0.9525. The results of the elevation accuracy verification (Fig. 6. Relationship between the inverted elevation values and the in situ elevation values. The green and cyan rectangles indicate outliers in the high and low tidal flats, respectively. The scatter within the red rectangle shows a high correlation.) show the inverted values display good correlation with the in situ elevation values. The scatter distribution matches the 1:1 line height, the R2 is 0.88, the RMSE is 0.30 m, and the MAE is 0.24 m. The accuracy verification results demonstrate this study can effectively extract tidal flats and accurately reflect their topographic information.

**Table 1**

**Results of the tidal flat spatial extent accuracy verification**

| Recognition type | Measured type | | Total rows | User accuracy/% | F1_score | Kappa |
|---|---|---|---|---|---|---|
| | Tidal flat | Other | | | | |
| Tidal flat | 341 | 13 | 354 | 96.33 | 0.9525 | 0.9069 |
| Other | 21 | 356 | 377 | 94.43 | | |
| Total columns | 362 | 369 | 731 | | | |
| Producer accuracy/% | 94.20 | 96.48 | | | | |
| Overall accuracy/% | 95.36 | | | | | |

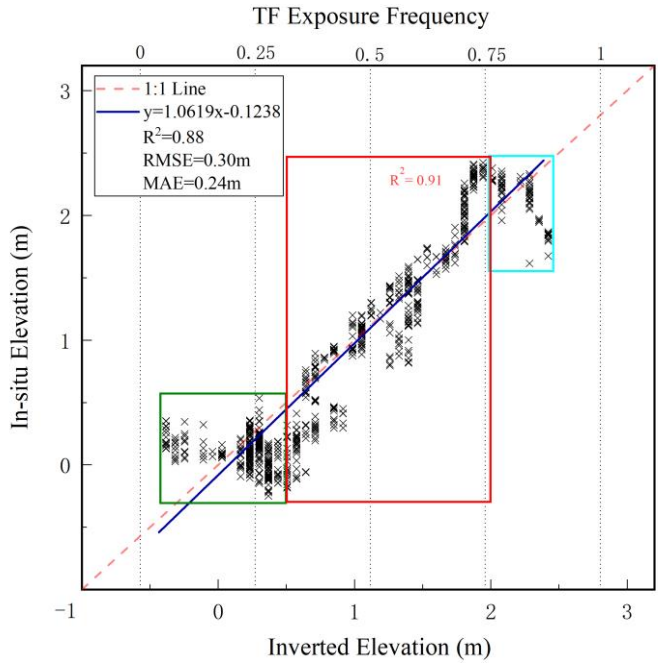

**Fig. 6.** Relationship between the inverted elevation values and the in situ elevation values. The green and cyan rectangles indicate outliers in the high and low tidal flats, respectively. The scatter within the red rectangle shows a high correlation.

3.3.2 Accuracy verification based on another set of ICESat-2 data

We conducted verification of the inversion results using another set of ICESat-2 data that passed through the study area's western region during a period of low tide (Fig. 7a). This area features extensive tidal flats, providing 13,362 ICESat-2 elevation points for validation, with a large sample size. After uniform scaling, 490 elevation values were included for validation of the accuracy. The scatter plot (Fig. 7b) shows a correlation coefficient ($R^2$) of 0.74 between the inverted values and the ICESat-2 elevation values, with a slope of 1.06.

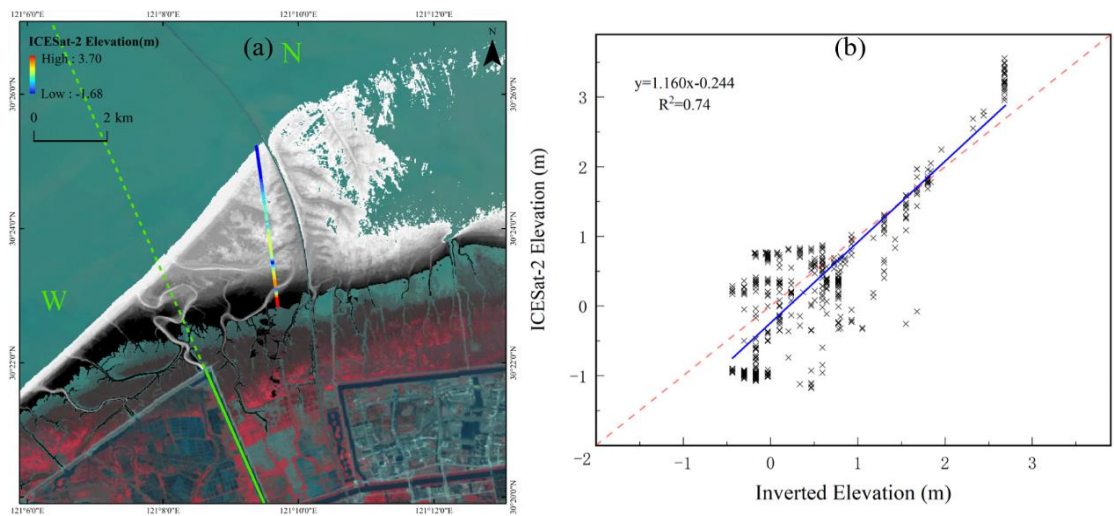

**Fig. 7.** (a) Location of the other set of ICESat-2 data. (b) Relationship between the inverted elevation values and the other set of ICESat-2 elevation values.

## 4. Discussion

4.1 Spatial distribution of tidal flats

The results of this study will be compared to two open-source tidal flat datasets, namely, UQD (Murray et al., 2019) (Fig. 8b) and CTF (Jia et al., 2021) (Fig. 8c). The UQD dataset is a global coastal zone tidal flat dataset developed by Murray et al., utilizing GEE and Landsat images and employing the random forest algorithm. The data can be accessed at https://www.intertidal.app/download. The CTF dataset is the Chinese tidal flat dataset by Jia et al. based on GEE and Sentinel images using the exponential extremum method, with data from the National Earth System Science Data Center, a national science and technology infrastructure platform (https://www.geodata.cn). The UQD dataset, in addition to classifying many dikes within non-permanent water as tidal flats but also classifies large supratidal areas as tidal flats. This may be because its random forest algorithm-based classification method does not consider tidal dynamic processes and considers bare land in the supratidal zone with spectral information expression to be similar to that of tidal flats, while areas covered by salt marsh vegetation in the supratidal zone are better segmented out. Certainly, the UQD dataset represents tidal flats from 2014 to 2016, which is a 5-year interval from the present study. During this period, the tidal flats were subject to natural sedimentation, erosion, and human activities, resulting in noticeable differences in both spatial distribution and topography compared to the current conditions. On the other hand, the CTF dataset represents the tidal flat area from 2019 to 2020. It also includes bare land in the supratidal zone as tidal flats. However, a significant number of areas that were classified as tidal flats in both this study and UQD were classified as permanent water in the CTF dataset. This discrepancy may be attributed to the use of a higher threshold value during the segmentation process, leading to an overestimation of the water areas.

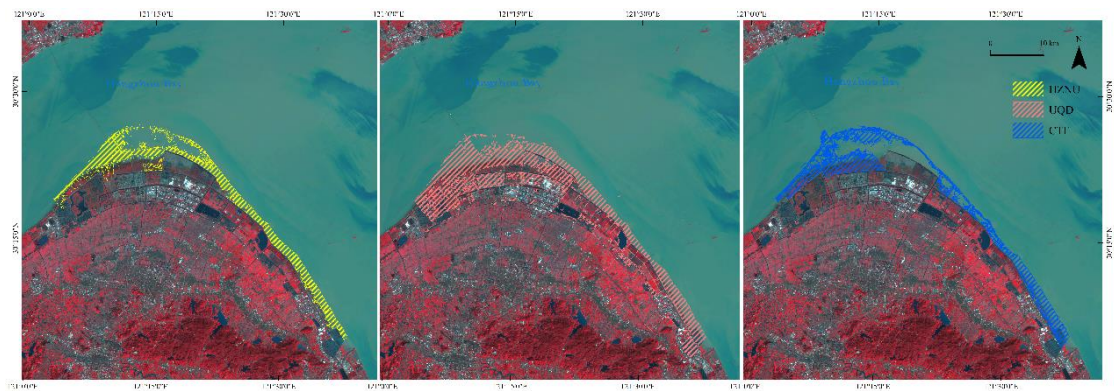

**Fig. 8.** (a) The distribution of tidal flats in this study, named HZNU. (b) UQD in Cixi City. (c) CTF in Cixi City.

4.2 Accuracy of tidal flat elevation inversion

4.2.1 Spatial heterogeneity

The accuracy of the elevation inversion results shows spatial heterogeneity, particularly in areas with lower tidal flat elevations where the inversion accuracy is generally lower.

The accuracy verification results based on UAV aerial photography show the inversion accuracy is relatively low for high tide flats and low tidal flats with inversion elevations below 0.5 m or above 2 m. However, the inversion performs better for medium tidal flats within the range of 0.5 to 2 m ($R^2$=0.91). In certain regions of the high tidal flats (above 2 m), the in situ elevations are significantly lower than the inverted elevations. The DEM imagery (Fig. 9a) reveals the presence of a mound-shaped feature at this site. However, due to the limited temporal resolution, the Sentinel-2 imagery (Fig. 9b) failed to capture the image when the mound was surrounded by inundation and its top was exposed. Consequently, the inversion result omitted this information. Furthermore, by comparing the ICESat-2 data used during inversion, it is noted the ICESat-2 data do not exhibit significant undulations in this area. It is speculated that this mound-shaped feature likely formed after the acquisition of the ICESat-2 data on March 21, 2021. As to why the tidal flats close to the edge of the dike are lower, we hypothesize that the dike acts as a solid impervious surface and that precipitation, as well as stagnant tidal water flowing down the dike, scours the tidal flats close to the dike.

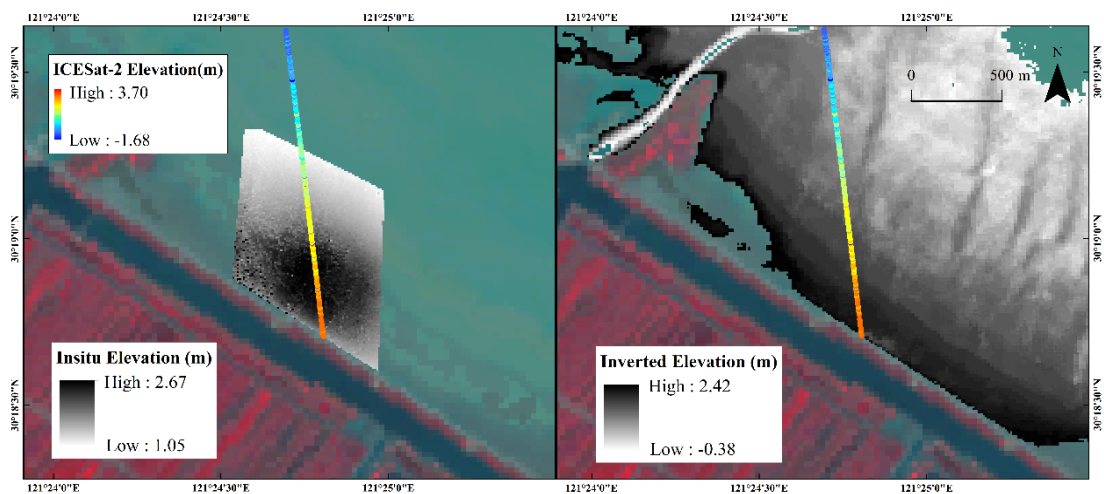

**Fig. 9.** Location of the outliers in the high tidal flat areas in **Fig. 6**. **Relationship between the inverted elevation values and the in situ elevation values. The green and cyan rectangles indicate outliers in the high and low tidal flats, respectively. The scatter within the red rectangle shows a high correlation.** (a) The DEM imagery (partial) revealing a mound-shaped feature. (b) The inversion result omitting this feature.

The accuracy verification results based on ICESat-2 data (Fig. 7b) demonstrate inversion accuracy is higher in high tidal flat areas than in low tidal flat areas. Since 75% of the validation samples are in the low tidal flat area, a significant amount of the micro-topographic information remains underrepresented, leading to a lower correlation within the low tidal flat area. Furthermore, topographic changes within the low tidal flat area over a span of two years may contribute to substantial differences in elevation when compared to the ICESat-2 validation data acquired on November 20, 2021.

4.2.2 Insufficient temporal resolution and more possibilities in modelling methods

The repeat period of Sentinel-2 data is 5 days, which already offers significant convenience for time-series monitoring. However, due to the cloudy and rainy meteorological characteristics in the subtropics, the number of effective observations remains inadequate. It is intuitive that denser data will enable a more detailed reflection of topographic information. In the context of this study, it can be said the temporal resolution is synonymous with the elevation resolution. In the future, we can explore the incorporation of additional data sources in our calculations. This requires careful consideration of how different datasets can be integrated. Moreover, we anticipate wider adoption and application of high-resolution imagery with shorter repeat periods. Building upon this foundation, we can then experiment with nonlinear or sub-regional modelling based on more precise tidal flat exposure frequencies.

4.3 Impacts of coastal engineering on tidal flats

It is evident that coastal construction has a significant influence on the development of adjacent tidal flats. In estuaries with high sediment content, the construction of dikes plays a crucial role in promoting sediment deposition on tidal flats (Mattheus et al., 2010; Shi et al., 2017). Hangzhou Bay serves as a typical estuary characterized by high sediment content. By effectively inverting the elevation of the tidal flats, the morphological evolution of the tidal flats can be analysed on a larger scale.

Studies on tidal flat morphodynamics suggest small waves, large tides, and sediment input contribute to the formation of convex tidal flat profiles, whereas large waves, small tides, and sediment output favour concave profiles (Bearman et al., 2010; Kirby, 2000). The development of tidal flats is a result of the complex interplay between hydrodynamic forces and sedimentation effects. In the absence of external interference, an "equilibrium profile" gradually emerges (Friedrichs, 2011). However, this equilibrium state is inherently unsustainable, as tidal flat development is constantly influenced by natural and anthropogenic factors. According to 1D sediment transport models proposed by Waeles et al. (2004), the predicted equilibrium profile initially exhibits convexity and transitions to concavity as wave energy increases.

Based on the tidal flat exposure frequency, we simulated the topography of the tidal flats in Sections E1 (Fig. 10a) and E2 (Fig. 10b). The tidal flats showed different morphologies in these two neighboring areas. Research conducted by Hu et al. (2019) reveals suspended sediment concentrations in the nearshore areas of E1 and E2 remain similar during neap, middle and spring tides. This indicates that Section E1 experiences larger waves and smaller tides, while the opposite holds true for Section E2. The construction of the spur dike further accentuates the differences in tidal flat development between these adjacent areas. In Section E2, the siltation-promoting effect of the dam is undoubtedly favorable. However, in Section E1, the tidal flats are high on three sides and low in the interior, showing a retreating profile, and the ideas in **Fig. 6** and 4.2.1 can be corroborated with this phenomenon. Some studies suggest construction of dikes can trigger both erosion (Bozek et al., 2005; Harmsworth et al., 1986) and siltation (Lotze et al., 2006; Mattheus et al., 2010) of tidal flats. However, it is relatively uncommon for two closely neighboring areas to exhibit such distinct differences. Due to this uncertainty, future polder construction may necessitate distinct strategies for these two regions.

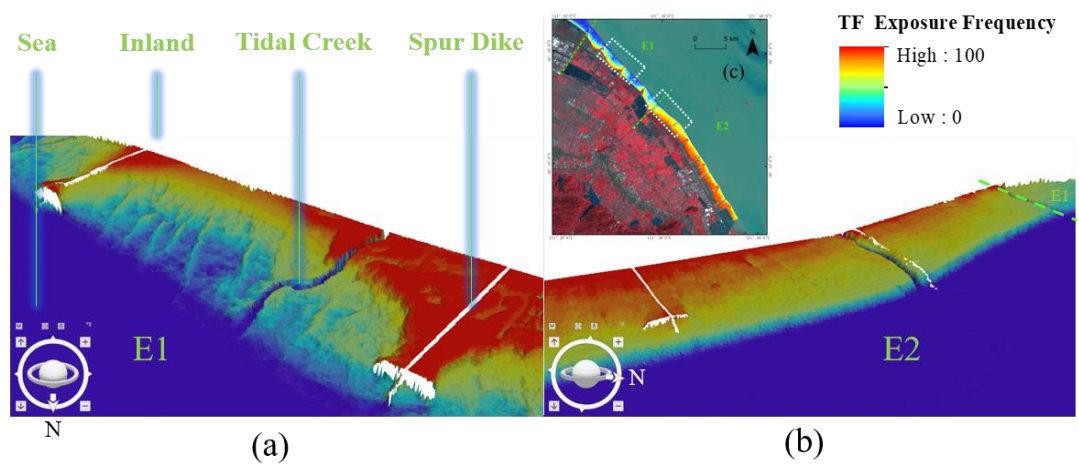

**Fig. 10.** Simulation of tidal flat topography based on the exposure frequency after 2.5x stretching. (a) Concave-up tidal flats in Section E1. (b) Convex-up tidal flats in Section E2. (c) Location of Sections E1 and E2.

**5. Conclusions**

This study introduces a highly automated method for mapping tidal flat topography based on the tidal flat exposure frequency. Using the Otsu algorithm in the GEE, the image pixels in the time-series Sentinel-2 MSI images are segmented into water and non-water pixels, and the frequency of non-water pixels at each pixel location, representing the tidal flat exposure frequency, is calculated. By constructing a 1D-linear regression model based on the tidal flat exposure frequency and the ICESat-2 elevation data, the elevation of the tidal flat can be inferred, demonstrating the potential of integrating multiple data sources for tidal flat monitoring. The main conclusions of the study are as follows:

1. The 1D-linear regression model constructed from tidal flat exposure frequency and ICESat-2 elevation data exhibits a strong positive correlation ($R^2=0.85$), indicating tidal flat exposure frequency can effectively represent topographical information of tidal flats.
2. The tidal flat area within Cixi City is 115.81 km$^2$. Through comparison and verification with multiple source image data, the method employed in this study demonstrates a high accuracy in determining the spatial extent of the tidal flats (F1_score=0.9525, Kappa=0.9069). In comparison to other tidal flat data, this study successfully removes the supratidal zone, which remains dry for extended periods, and provides a comprehensive representation of important features such as tidal channel morphology.
3. The elevation range of the tidal flats in the study area is between -0.42 and 2.73 m. The validity of the inversion results was confirmed through verification with photogrammetric elevation information, showing a strongly positive correlation ($R^2=0.88$). Additionally, the validation based on over ten thousand ICESat-2 data points also demonstrates a favourably positive correlation ($R^2=0.74$).

On the one hand, this method offers a reliable means to acquire spatial and elevation information of tidal flats, which holds significant implications for the dynamic monitoring and management of tidal resources. It also serves as a valuable data foundation for studies on tidal flat morphodynamics and carbon sink estimation. On the other hand, the calculation of tidal flat exposure frequency in this study is seamlessly implemented using the GEE, benefiting from easy accessibility and widespread distribution of ICESat-2 data. By simply adjusting the region of interest (ROI) within the GEE, this inversion method can be easily replicated over a global majority of regions, enabling large-scale mapping of tidal flat topography. In summary, we believe with the support of cloud computing and globally available data, that this method holds vast application prospects. However, we must acknowledge that tidal flats, being a highly dynamic landscape with intermittent appearances, still face limitations in the capturing of detailed topographic information due to the relatively low temporal resolution of the employed time-series approach. With the integration of more densely populated data sources in the future, we will incorporate diverse tidal flat morphologies and utilize different modelling methods across various study areas. This will further enhance the reliability of the method.

Declaration of Competing Interest
　　The authors declare that they have no known competing financial interests or personal relationships that could have appeared to influence the work reported in this paper.

Acknowledgements
　　This research was supported by the Global Change and Air-Sea Interaction Project of China

(Grants #JC-YGFW-YGJZ). We acknowledge the NASA Earth Data portal for providing the ICESat-2 data and the Resource and Environment Science and Data Center of the Chinese Academy of Sciences for the administrative boundary data, used in this study.

References

Allen, G. H., Pavelsky, T. M., 2018. Global extent of rivers and streams. *Science*, 361(6402), 585-588. https://doi.org/10.1126/science.aat0636.

Bearman, J. A., Friedrichs, C. T., Jaffe, B. E., et al., 2010. Spatial trends in tidal flat shape and associated environmental parameters in South San Francisco Bay. *J. Coastal Res*, 26(2), 342-349. https://doi.org/10.2112/08-1094.1.

Bozek, C. M., Burdick, D. M., 2005. Impacts of seawalls on saltmarsh plant communities in the Great Bay Estuary, New Hampshire USA. *Wetl Ecol Manag*, 13, 553-568. https://doi.org/10.1007/s11273-004-5543-z.

Chang, M., Li, P., Li, Z., et al., 2022. Mapping Tidal Flats of the Bohai and Yellow Seas Using Time Series Sentinel-2 Images and Google Earth Engine. *Remote Sens.*, 14(8), 1789. https://doi.org/10.3390/rs14081789.

Einsele, G., Yan, J., Hinderer, M., 2001. Atmospheric carbon burial in modern lake basins and its significance for the global carbon budget. *Glob Planet Change*, 30(3), 167-195. https://doi.org/10.1016/S0921-8181(01)00105-9.

Feyisa, G. L., Meilby, H., Fensholt, R., et al., 2014. Automated Water Extraction Index: A new technique for surface water mapping using Landsat imagery. *Remote Sens Environ*, 140, 23-35. https://doi.org/10.1016/j.rse.2013.08.029.

Friedrichs, C. T., 2011. 3.06 - Tidal Flat Morphodynamics: A Synthesis. *Treatise on Estuarine and Coastal Science*, 130-170. https://doi.org/10.1016/B978-0-12-374711-2.00307-7.

Gorman, L., Morang, A., Larson, R., 1998. Monitoring the coastal environment; Part IV: Mapping, shoreline changes, and bathymetric analysis. *J. Coastal Res*, 14(1), 61–92. http://www.jstor.org/stable/4298762 61-92.

Harmsworth, G., Long, S., 1986. An assessment of saltmarsh erosion in Essex, England, with reference to the Dengie Peninsula. *Biol. Conserv.*, 35(4), 377-387. https://doi.org/10.1016/0006-3207(86)90095-9.

Healy, T., Wang, Y., Healy, J.-A. (Eds.), 2002. Muddy coasts of the world: processes, deposits and function. Elsevier, Amsterdam.

Hu, Y., Yu, Z., Zhou, B., et al., 2019. Tidal-driven variation of suspended sediment in Hangzhou Bay based on GOCI data. *Int J Appl Earth Obs Geoinf*, 82, 101920. https://doi.org/10.1016/j.jag.2019.101920.

James, T., Schillaci, C., Lipani, A., 2021. Convolutional neural networks for water segmentation using sentinel-2 red, green, blue (RGB) composites and derived spectral indices. *Int J Remote Sens*, 42(14), 5338-5365. https://doi.org/10.1080/01431161.2021.1913298.

Ji, L., Zhang, L., Wylie, B., 2009. Analysis of dynamic thresholds for the normalized difference water index. *Photogramm Eng Remote Sensing*, 75(11), 1307-1317. https://doi.org/10.14358/PERS.75.11.1307.

Jia, M., Wang, Z., Mao, D., et al., 2021. Rapid, robust, and automated mapping of tidal flats in China using time series Sentinel-2 images and Google Earth Engine. *Remote Sens Environ,* 255, 112285. https://doi.org/10.1016/j.rse.2021.112285.

Kalacska, M., Chmura, G., Lucanus, O., Bérubé, D., Arroyo-Mora, J., 2017. Structure from motion will revolutionize analyses of tidal wetland landscapes. *Remote Sens Environ*, 199, 14-24. https://doi.org/10.1016/j.rse.2017.06.023.

Kim, J.-W., Lu, Z., Jones, J. W., et al., 2014. Monitoring Everglades freshwater marsh water level using L-band synthetic aperture radar backscatter. *Remote Sens Environ*, 150, 66-81. https://doi.org/10.1016/j.rse.2014.03.031.

Kirby, R., 2000. Practical implications of tidal flat shape. *Cont Shelf Res*, 20(10-11), 1061-1077. https://doi.org/10.1016/S0278-4343(00)00012-1.

Li, H., Cutler, M., Zhang, D., et al., 2022a. Retrieval of Tidal Flat Elevation Based on Remotely Sensed Moisture Approach. IEEE J Sel Top Appl Earth Obs Remote Sens, 15, 5357-5370. https://doi.org/10.1109/JSTARS.2022.3187148.

Li, Y., Niu, Z., 2022b. Systematic method for mapping fine-resolution water cover types in China based on time series Sentinel-1 and 2 images. *Int J Appl Earth Obs Geoinf*, 106, 102656. https://doi.org/10.1016/j.jag.2021.102656.

Lotze, H. K., Lenihan, H. S., Bourque, B. J., et al., 2006. Depletion, degradation, and recovery potential of estuaries and coastal seas. *Science*, 312(5781), 1806-1809. https://doi.org/10.1126/science.1128035.

Luo, S., Wang, C., Xi, X., et al., 2017. Retrieving aboveground biomass of wetland Phragmites australis (common reed) using a combination of airborne discrete-return LiDAR and hyperspectral data. *Int J Appl Earth Obs Geoinf*, 58, 107-117. https://doi.org/10.1016/j.jag.2017.01.016.

Markus, T., Neumann, T., Martino, A., et al., 2017. The Ice, Cloud, and land Elevation Satellite-2 (ICESat-2): science requirements, concept, and implementation. *Remote Sens Environ*, 190, 260-273. https://doi.org/10.1016/j.rse.2016.12.029.

Mason, D., Davenport, I., Flather, R., et al., 2001. A sensitivity analysis of the waterline method of constructing a digital elevation model for intertidal areas in ERS SAR scene of eastern England. *Estuarine, Estuar. Coast. Shelf Sci.*, 53(6), 759-778. https://doi.org/10.1006/ecss.2000.0789.

Mattheus, C. R., Rodriguez, A. B., McKee, B. A., et al., 2010. Impact of land-use change and hard structures on the evolution of fringing marsh shorelines. *Estuar. Coast. Shelf Sci.,* 88(3), 365-376. https://doi.org/10.1016/j.ecss.2010.04.016.

McFeeters, S. K., 1996. The use of the Normalized Difference Water Index (NDWI) in the delineation of open water features. *Int J Remote Sens*, 17(7), 1425-1432. https://doi.org/10.1080/01431169608948714.

Murray, N. J., Phinn, S. R., DeWitt, M., et al.,2019. The global distribution and trajectory of tidal flats. *Nature*, 565(7738), 222-225. https://doi.org/10.1038/s41586-018-0805-8.

Neumann, T., Brenner, A., Hancock, D., et al., 2021. ATLAS/ICESat-2 L2A Global Geolocated Photon Data, Version 5. Boulder, CO: NASA National Snow and Ice Data Center Distributed Active Archive Center. https://doi.org/10.5067/ATLAS/ATL03.005.

Otsu, N., 1979. A threshold selection method from gray-level histograms. *IEEE Trans. Syst. Man Cybern. Syst.*, 9(1), 62-66. http://doi.org/10.1109/TSMC.1979.4310076.

Parrish, C. E., Magruder, L. A., Neuenschwander, A. L., et al., 2019. Validation of ICESat-2 ATLAS bathymetry and analysis of ATLAS's bathymetric mapping performance. *Remote Sens.*, 11(14), 1634. https://doi.org/10.3390/rs11141634.

Pekel, J.-F., Cottam, A., Gorelick, N., et al., 2016. High-resolution mapping of global surface water and its long-term changes. *Nature*, 540(7633), 418-422. https://doi.org/10.1038/nature20584.

Reise, K., Baptist, M., Burbridge, P., et al., 2010. The Wadden Sea a universally outstanding tidal wetland. *Wadden Sea Ecosystem*. 29, 7-24. https://www.waddensea-worldheritage.org/sites/default/files/2010_Ecosystem29_the%20wadden%20sea%202010.pdf#page=9.

Shi, B., Yang, S., Wang, Y., et al., 2017. Role of wind in erosion-accretion cycles on an estuarine mudflat. *J. Geophys. Res. Oceans*, *122*(1), 193-206. https://doi.org/10.1002/2016JC011902.

Waeles, B. t., Le Hir, P., Jacinto, R. S., 2004. Modélisation morphodynamique cross-shore d'un estran vaseux. *CR GEOSCI*, 336(11), 1025-1033. https://doi.org/10.1016/j.crte.2004.03.011.

Wang, X., Xie, S., Zhang, X., et al., 2018. A robust Multi-Band Water Index (MBWI) for automated extraction of surface water from Landsat 8 OLI imagery. *Int J Appl Earth Obs Geoinf*, 68, 73-91. https://doi.org/10.1016/j.jag.2018.01.018.

Xie, C., Xu, J., Shao, Y., et al., 2015. Long term detection of water depth changes of coastal wetlands in the Yellow River Delta based on distributed scatterer interferometry. *Remote Sens Environ*, 164, 238-253. https://doi.org/10.1016/j.rse.2015.04.010.

Xu, H., 2006. Modification of normalised difference water index (NDWI) to enhance open water features in remotely sensed imagery. *Int J Remote Sens*, 27(14), 3025-3033. https://doi.org/10.1080/01431160600589179.